\documentclass[aps, prl, twocolumn, floatfix, superscriptaddress]{revtex4-1} 
\usepackage{graphicx}
\usepackage{epstopdf}
\usepackage{bm, amsmath, amssymb}
\usepackage{float}
\usepackage{placeins}
\usepackage{color}

\newcommand{\ket}[1]{\left|#1\right>}
\newcommand{\bra}[1]{\left<#1\right|}

\begin{document}

\title{Continuous quantum nondemolition measurement of the transverse component of a qubit}

\author{U. Vool}
\email{uri.vool@yale.edu}
\affiliation{Department of Applied Physics and Physics, Yale University, New Haven, CT 06520}
\author{S. Shankar}
\affiliation{Department of Applied Physics and Physics, Yale University, New Haven, CT 06520}
\author{S. O. Mundhada}
\affiliation{Department of Applied Physics and Physics, Yale University, New Haven, CT 06520}
\author{N. Ofek}
\affiliation{Department of Applied Physics and Physics, Yale University, New Haven, CT 06520}
\author{A. Narla}
\affiliation{Department of Applied Physics and Physics, Yale University, New Haven, CT 06520}
\author{K. Sliwa}
\affiliation{Department of Applied Physics and Physics, Yale University, New Haven, CT 06520}
\author{E. Zalys-Geller}
\affiliation{Department of Applied Physics and Physics, Yale University, New Haven, CT 06520}
\author{Y. Liu}
\affiliation{Department of Applied Physics and Physics, Yale University, New Haven, CT 06520}
\author{L. Frunzio}
\affiliation{Department of Applied Physics and Physics, Yale University, New Haven, CT 06520}
\author{R. J. Schoelkopf}
\affiliation{Department of Applied Physics and Physics, Yale University, New Haven, CT 06520}
\author{S. M. Girvin}
\affiliation{Department of Applied Physics and Physics, Yale University, New Haven, CT 06520}
\author{M. H. Devoret}
\affiliation{Department of Applied Physics and Physics, Yale University, New Haven, CT 06520}

\date{\today}

\begin{abstract}

Quantum jumps of a qubit are usually observed between its energy eigenstates, also known as its longitudinal pseudo-spin component. Is it possible, instead, to observe quantum jumps between the transverse superpositions of these eigenstates? We answer positively by presenting the first continuous quantum nondemolition measurement of the transverse component of an individual qubit. In a circuit QED system irradiated by two pump tones, we engineer an effective Hamiltonian whose eigenstates are the transverse qubit states, and a dispersive measurement of the corresponding operator. Such transverse component measurements are a useful tool in the driven-dissipative operation engineering toolbox, which is central to quantum simulation and quantum error correction.

\end{abstract}

\maketitle

A qubit is traditionally described in the computational basis, formed by its energy (longitudinal) eigenstates, the ground state  $\ket{g}$ and the excited state $\ket{e}$.
It is possible to continuously monitor~\cite{Bergquist1986,Sauter1986,Gleyzes2007,Vijay2011}  the longitudinal component of the qubit, namely the operator $\ket{e}\bra{e} - \ket{g}\bra{g}$, which corresponds to $\bm{\sigma_z}$ in the Pauli matrix basis~\cite{NielsenChuang}. However, no experiment has yet continuously monitored the transverse component of an individual qubit, i.e. the operator $\ket{g}\bra{e} + \ket{e}\bra{g}$, equivalent to $\bm{\sigma_x}$, even though in the seminal field of magnetic resonance, it is the transverse component of an ensemble of spins which is the most common measurement~\cite{Abragam1961,Slichter1990}.
The difficulty of continuous measurement of $\bm{\sigma_x}$ arises because it does not commute with the static Hamiltonian ($\propto \bm{\sigma_z}$), and hence the projected states evolve during the measurement process. To overcome this difficulty, we would like our measurement axis to evolve according to the Hamiltonian, and appear as a $\bm{\sigma_x}$ measurement in the rotating frame in which the static Hamiltonian has been suppressed.
Of course, it is possible to stroboscopically measure any component of the pseudo-spin in the Larmor frame, but in this Letter we are focusing on continuous, rather than discrete projective measurements of $\bm{\sigma_x}$.

Why would one want to perform a continuous $\bm{\sigma_x}$ measurement of a qubit? One direct application is to probe decoherence mechanisms with higher frequency and time resolution than usual stroboscopic methods~\cite{Biercuk2009,Bylander2011,Sank2012,Riste2013a}. More fundamentally, however, we would like to master the methods by which we impose not only a given field belonging to the Hilbert space of the system (here $\bm{\sigma_x}$ rather than $\bm{\sigma_z}$), but also to implement the dissipation conjugate to that field and the measurement of its associated fluctuations. This mastery is particularly timely in view of recent progress in quantum control and bath engineering~\cite{Haroche2006,Wiseman2009,Poyatos1996,Krauter2011,Vijay2012, Murch2012,Campagne-Ibarcq2013,Riste2013,Lin2013,Kerckhoff2013,Shankar2013,Nigg2014,Leghtas2015} that allow us to synthesize desired Hamiltonian and open-system operations. Moreover, such techniques are at the heart of quantum error correction~\cite{Chiaverini2004,Reed2012,Waldherr2014,Kelly2015,Corcoles2015}. 

In this Letter, we implement a quantum nondemolition (QND) measurement of the transverse component of a superconducting artificial atom, embedded in a traditional circuit-QED setup~\cite{Blais2004}. Using pump tones, we engineer a new effective qubit in the transverse basis, dispersively coupled to a cavity mode. This new Hamiltonian commutes with the desired measurement operator $\bm{\sigma_x}$. Thus, with the help of a quantum-limited amplifier~\cite{Castellanos-Beltran2008,Bergeal2010,Hatridge2011}, we perform continuous, projective measurements and observe quantum jumps~\cite{Vijay2011} between the eigenstates of the effective qubit, which are the transverse superpositions of the original qubit eigenstates. 

Our system consists of a transmon qubit~\cite{Koch2007} with frequency $\omega_q$ coupled to a 3D superconducting cavity~\cite{Paik2011} with frequency $\omega_c$ (see Fig.~1a). We apply two pump tones to the system, a sideband tone detuned from the cavity frequency by $\Delta_c$ and a Rabi tone at the qubit frequency (see Fig.~1b - the additional readout tone is a probe that will be described later). Treating the transmon qubit as a 2-level system and within the dispersive approximation, we can write the Hamiltonian for our system:
\begin{equation}
\begin{split}
\bm{H}/\hbar&=\omega_c \bm{a^{\dagger}a} + \frac{\omega_q}{2} \bm{\sigma_{z}} + \frac{\chi}{2}\bm{a^{\dagger}a}\bm{\sigma_z} \\
&+ \Omega_\mathrm{R} \cos(\omega_q t)\bm{\sigma_x} + \epsilon_{sb} \cos (\omega_{sb} t)(\bm{a}+\bm{a^\dagger})
\end{split}
\end{equation}
where $\bm{a}$ is the cavity lowering (annihilation) operator and $\bm{\sigma_{z}}$ is the energy operator of the qubit, modeled here as a two-level system. $\chi$ is the dispersive shift between the qubit and cavity, $\Omega_\mathrm{R}$ is the amplitude of the Rabi drive in frequency units (Rabi frequency), and $\epsilon_{sb}$ is the amplitude of the sideband drive at frequency $\omega_{sb} = \omega_c-\Delta_c$. 
By moving to a rotated and displaced frame (see~\cite{Supplementary} for details), we can approximate our system by the time-independent Hamiltonian:
\begin{equation}
\label{Jaynessigx}
\bm{H_{\mathrm{JC}}}/\hbar=\Delta_c \bm{d^{\dagger}d}  + \frac{\Omega_\mathrm{R}}{2} \bm{\sigma_x}+ \frac{\chi}{2}(\bar{a}^\ast \bm{d\sigma^+_x} + \bar{a} \bm{d^{\dagger} \sigma^-_x}),
\end{equation}
where we introduce the displaced cavity lowering operator $\bm{d} = \bm{a} - \bar{a}$ and the steady state amplitude of the cavity $\bar{a} = \frac{-\epsilon_{sb}}{\Delta_c -i \kappa/2}$, where $\kappa$ is the cavity decay rate (see Fig.~1b).
Notice the qubit now has an energy splitting in the $\bm{\sigma_x}$ basis given by $\Omega_\mathrm{R}$ and we have introduced the raising (lowering) operators $\bm{\sigma^\pm_x}=(\bm{\sigma_z} \mp i \bm{\sigma_y})/2$  between the eigenstates of $\bm{\sigma_x}$, which obey $\bm{\sigma_z} = \bm{\sigma^+_x} + \bm{\sigma^-_x}$. 

This Hamiltonian resembles the Jaynes-Cummings ~\cite{Jaynes1963,Haroche2006} (JC) Hamiltonian $\omega^\mathrm{eff}_c \bm{a^\dagger a} + \frac{\omega^\mathrm{eff}_q}{2}\bm{\sigma_z} + g_\mathrm{eff}(\bm{a \sigma^+}+\bm{a^\dagger \sigma^-})$, where the effective cavity is the displaced cavity, the effective atom is the $\bm{\sigma_x}$ qubit and the coupling between them is $g_\mathrm{eff} = \frac{\chi |\bar{a}|}{2}$. Thus, we have created an effective JC Hamiltonian where both the frequencies and the coupling term are completely tunable in situ by varying the amplitudes and frequencies of the Rabi and sideband pumps (see also Ref.~\onlinecite{Ballester2012} for a different method to generate an effective JC Hamiltonian). 

Since the parameters are completely tunable, we can operate in either the strong ($g_\mathrm{eff} > \kappa$) or weak ($g_\mathrm{eff} < \kappa$) coupling regime, and also in the resonant ($|\Omega_\mathrm{R} - \Delta_c| \ll g_\mathrm{eff}$) or dispersive ($|\Omega_\mathrm{R} - \Delta_c| \gg g_\mathrm{eff}$) regime.
In the resonant regime, the scheme becomes identical to that used in Ref.~\onlinecite{Murch2012} to cool a superconducting qubit to an eigenstate of $\bm{\sigma_x}$. This process can be interpreted as a sideband cooling~\cite{Diedrich1989,Hamann1998,Teufel2011} scheme where a red-detuned drive cools the effective qubit, utilizing cavity dissipation. In our JC picture, the fast decay rate ($\kappa$) of the cavity simply takes the system to its ground state, in which the qubit is in $\ket{-}$ - an eigenstate of $\bm{\sigma_x}$.

This Letter will instead focus on the dispersive regime. In the limit $|\Omega_\mathrm{R} - \Delta_c| \gg \frac{\chi |\bar{a}|}{2}$,  the interaction becomes dispersive and we can diagonalize the system~\cite{Schrieffer1966,Blais2004,Supplementary} to obtain the Hamiltonian:
\begin{equation}
\label{dispsigx}
\bm{H_{disp}}/\hbar=\Delta_c \bm{d^{\dagger}d} + \frac{\Omega_\mathrm{R}+\zeta/2}{2} \bm{\sigma_x}+ \frac{\zeta}{2}\bm{d^\dagger d \sigma_x},
\end{equation}
where $\zeta \simeq g_\mathrm{eff}^2/(\Omega_\mathrm{R} - \Delta_c)$ is a dispersive shift between the cavity and the $\bm{\sigma_x}$ qubit. Note that unlike the resonant sideband cooling regime, the existence of the dispersive shift has no opto-mechanical analog~\cite{Aspelmeyer2014} as it relies on the nonlinearity of the system. Another novelty of Eq. \ref{dispsigx} is that $\zeta$ is tunable in situ. The absence of a frequency matching condition in this regime means that fine-tuning of the pump tones is not required. Thus, we can perform a continuous QND measurement of the $\bm{\sigma_x}$ component of the qubit using traditional dispersive readout protocols~\cite{Blais2004}. This is done by applying a readout tone at $\Delta_c$ in the effective frame, or at $\omega_c$ in the lab frame (see Fig.~1b).

\begin{figure} [!t]
 \includegraphics[angle = 0, width = \columnwidth]{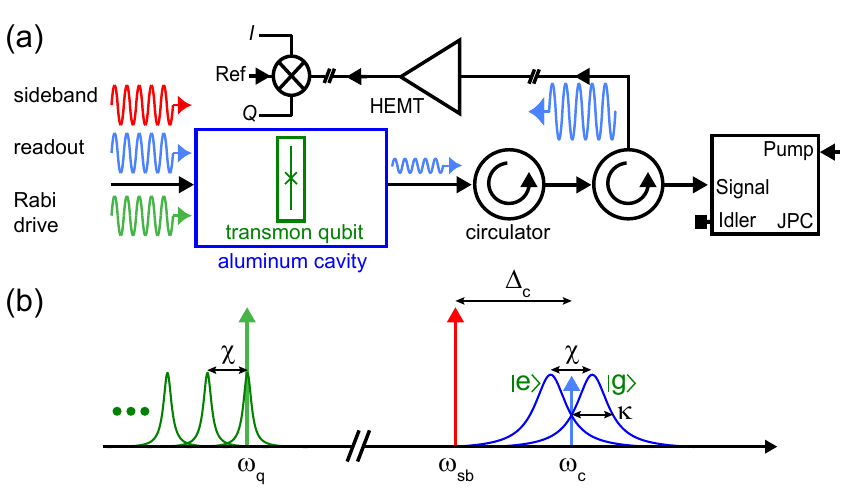}
 \caption{\label{fig1}  (a) Schematic of the experiment setup. A 3D transmon qubit - cavity system is continuously driven by 3 tones termed as sideband (red), readout (blue) and Rabi (green). The readout tone is transmitted through the output port, amplified by a Josephson Parametric Converter amplifier and demodulated at room temperature to give $I, Q$ signals. (b) Frequency landscape. Our system consists of a qubit at frequency $\omega_q$ and cavity with qubit-state-dependent frequency $\omega_c\pm\chi/2$ and linewidth $\kappa$. The qubit frequency correspondingly depends on the number of photons in the cavity, changing by $\chi$ for every photon. We apply a strong sideband tone (red) detuned from the cavity frequency by $\Delta_c$ and a strong Rabi tone (green) at the qubit frequency. Readout is performed by applying a weak readout tone (blue) at the cavity resonance frequency $\omega_c$ to readout the system. 
 }
 \end{figure}

Our experimental setup consists of a transmon qubit at frequency $\omega_q /2\pi = 4.9\:\mathrm{GHz}$ coupled to a 3D aluminum cavity with frequency $\omega_c /2\pi = 7.48\:\mathrm{GHz}$. The cavity has two coupling pins, a weakly coupled input pin with quality factor $Q_{in} \simeq 5\cdot 10^5$ and strongly coupled output port with $Q_{out} \simeq 1900$, and thus a total decay rate $\kappa /2\pi = 4\:\mathrm{MHz}$. The qubit has a decay time $T_1 = 90\:\mathrm{\mu s}$, a coherence time of  $T_{2R} = 40\:\mathrm{\mu s}$, and in thermal equilibrium it has an excited state population of $12\%$. The dispersive coupling between the qubit and cavity is $\chi /2\pi = -3.2\:\mathrm{MHz}$. All drives were applied at the input port and the signal from the output port was amplified by a Josephson Parametric Converter (JPC), a nearly quantum-limited amplifier~\cite{Bergeal2010, Abdo2013}, before being demodulated to extract $I, Q$ quadrature measurement outcomes.

We applied a cavity sideband tone with $\Delta_c/2\pi = 15\:\mathrm{MHz}$ such that the cavity steady state population was $\bar{n}_{sb} = |\bar{a}|^2= 12$ photons, which set $g_\mathrm{eff}/2\pi = \chi |\bar{a}|/4\pi = 5.5\:\mathrm{MHz}$. We also applied a resonant readout tone on the cavity such that $\bar{n} = 0.9$ photons. The histograms in Fig.~2 correspond to a $1\:\mathrm{s}$ long readout pulse demodulated every $400\:\mathrm{ns}$. In Fig.~2a the Rabi tone was off and our measurement projects the system to eigenstates of $\bm{\sigma_z}$, the top (bottom) distributions corresponding to the ground (excited) states of the qubit.  As we turn up the Rabi tone, our measurement no longer commutes with the system Hamiltonian and a ``competition'' takes place between the measurement and Rabi drive, sometimes called the quantum Zeno effect{~\cite{Misra1977, Gambetta2008}. When the Rabi frequency is below our measurement rate~\cite{Gambetta2008} $\Gamma_m/2\pi = 2.8\:\mathrm{MHz}$ we can still observe two distinct states (Fig.~2b), but as it gets much stronger, the measurement can no longer distinguish them (Fig.~2c). 

\begin{figure*} [!t]
\includegraphics[angle = 0, width = \textwidth]{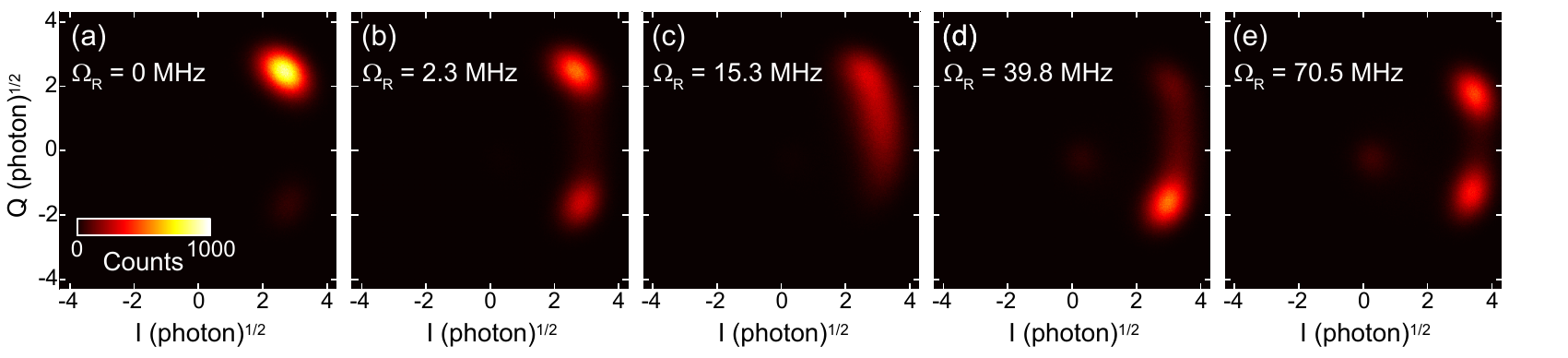}
\caption{\label{fig2}   Histograms of $I, Q$ measurements in the presence of the readout, sideband and Rabi drives for various indicated values of $\Omega_\mathrm{R}$. For each $\Omega_\mathrm{R}$, a $1\:\mathrm{s}$ continuous trace was recorded and integrated in $400\:\mathrm{ns}$ chunks to give an $I,Q$ value, in units of $\sqrt{\mathrm{photon}}$ in the integrated chunk. The sideband and readout drive powers resulted in a steady state population of $\bar{n}_{sb} = 12$ and $\bar{n} = 0.9$ photons in the cavity respectively. The two gaussian distributions corresponding to the eigenstates of $\bm{\sigma_z}$ are visible in (a) and disappear as $\Omega_\mathrm{R}$ is increased. For large values of $\Omega_\mathrm{R}$ two new distributions appear, corresponding to the eigenstates of $\bm{\sigma_x}$. The faint distribution near the center in (e) corresponds to the 2nd excited state of the qubit ($\ket{f}$).
}
\end{figure*}

However, as the Rabi frequency increases beyond $\Delta_c$, two distinct distributions reappear (Fig.~2d,e). Fig.~2d shows the formation of two distributions, with most of the population in the bottom one and a faint top distribution. As the Rabi frequency increases even further, the bimodality of the histogram becomes more marked and the two sub-populations become more equal (Fig.~2e). We understand this behavior as follows: when the system in Eq. \ref{Jaynessigx} is still close to the resonant regime $g_\mathrm{eff} \sim |\Omega_\mathrm{R}-\Delta_c|$ not quite in the dispersive limit, cavity dissipation cools the system to $\ket{-} = (\ket{g}-\ket{e})/\sqrt{2}$, analogous to the Purcell effect~\cite{Purcell1946, Houck2008, Murch2012, Supplementary}. Fig.~2d shows this effect, with the bottom distribution corresponding to the ``ground state'' $\ket{-}$ and the faint top distribution corresponding to $\ket{+}=(\ket{g}+\ket{e})/\sqrt{2}$. As the Rabi frequency increases, the coupling becomes more dispersive ($g_\mathrm{eff} \ll |\Omega_\mathrm{R}-\Delta_c|$) and the cooling effect weakens, as shown by the relative populations in Fig.~2e. The dispersive coupling parameter $\zeta$ can be extracted from these measurements and agrees with our theoretical prediction~\cite{Supplementary}. Notice that as we increased the Rabi frequency, a faint distribution appeared near the center of the figure. This distribution corresponds to the 2nd excited ($\ket{f}$) state of the qubit, which has an $8\%$ population in Fig.~2e.

To prove that this measurement projects the qubit to the eigenstates of $\bm{\sigma_x}$, we prepared a well-defined state before performing our measurement. We prepared the ground state $\ket{g}$ by standard $\bm{\sigma_z}$ dispersive measurement, applied a pulse to prepare a state on the Bloch sphere, and then turned on the Rabi tone with $\Omega_\mathrm{R}/2\pi = 70\: \mathrm{MHz}$ and the readout tone (the sideband tone is applied throughout the experiment to maintain the same qubit frame - since the presence of this tone causes a Stark shift of the qubit frequency). In Fig.~3a(b), we prepared the system in $\ket{-}$ ($\ket{+}$) and observed 90$\%$ (85$\%$) population in the bottom (top) distribution, limited by the state lifetime and the quality of our pulses~\cite{Supplementary}. In Fig.~3c,d, we prepared the qubit in $\ket{g}$ and $\ket{i} = (\ket{g}+i\ket{e})/\sqrt{2}$ respectively, and observed a nearly 50:50 population in both distributions. The separation between the two states is 5.4 $\sigma$, such that the infidelity due to the noise in the measurement chain is only about $1\%$.

The characterization of the measurement (in particular its fidelity) is further examined by rotating the qubit about the 3 main axes of the Bloch sphere (Fig.~3e-g). For each state we measure the expectation value $\left<\bm{\sigma_x}\right>$. The dashed red lines correspond to the expected result of an ideal measurement of $\left<\bm{\sigma_x}\right>$. The solid red lines are given by a simulation of the open qubit-cavity system to model state preparation errors, along with a scaling of $88\%$ and a shift of $2\%$ due to measurement imperfection~\cite{Supplementary}.
In Fig.~3e,f, we observed the expected sinusoidal behavior as we project unto eigenstates of $\bm{\sigma_x}$. Fig.~3g shows the axis perpendicular to $\bm{\sigma_x}$ and so should have a constant expectation $\left<\bm{\sigma_x}\right> = 0$. We observed a $0.2$ deviation from this distribution, leaning towards $\ket{-}$ ($\ket{+}$) for negative (positive) angles.
This deviation, also captured by our theoretical prediction, is an artifact of the state preparation, albeit an interesting one as it is also an effect of the $\bm{\sigma_x}$ coupling. During the preparation pulse for states on the $\bm{\sigma_y}-\bm{\sigma_z}$ plane, there is a Rabi drive along the $\bm{\sigma_x}$ axis, and so with the help of the sideband tone the system is cooled to its lower eigenstate - which changes depending on the direction of the pulse. 
Fig.~3e shows good agreement between the data and the theoretical prediction. In Fig.~3f,g, the discrepancy can be attributed to a slight nonlinearity in the relation between the amplitude of the preparation pulse and the Bloch sphere angle. From these experiments, we conclude that the average fidelity of our $\bm{\sigma_x}$ measurement is $88\%$, which agrees with our theoretical prediction~\cite{Supplementary}.

\begin{figure} [!t]

\includegraphics[angle = 0, width = \columnwidth]{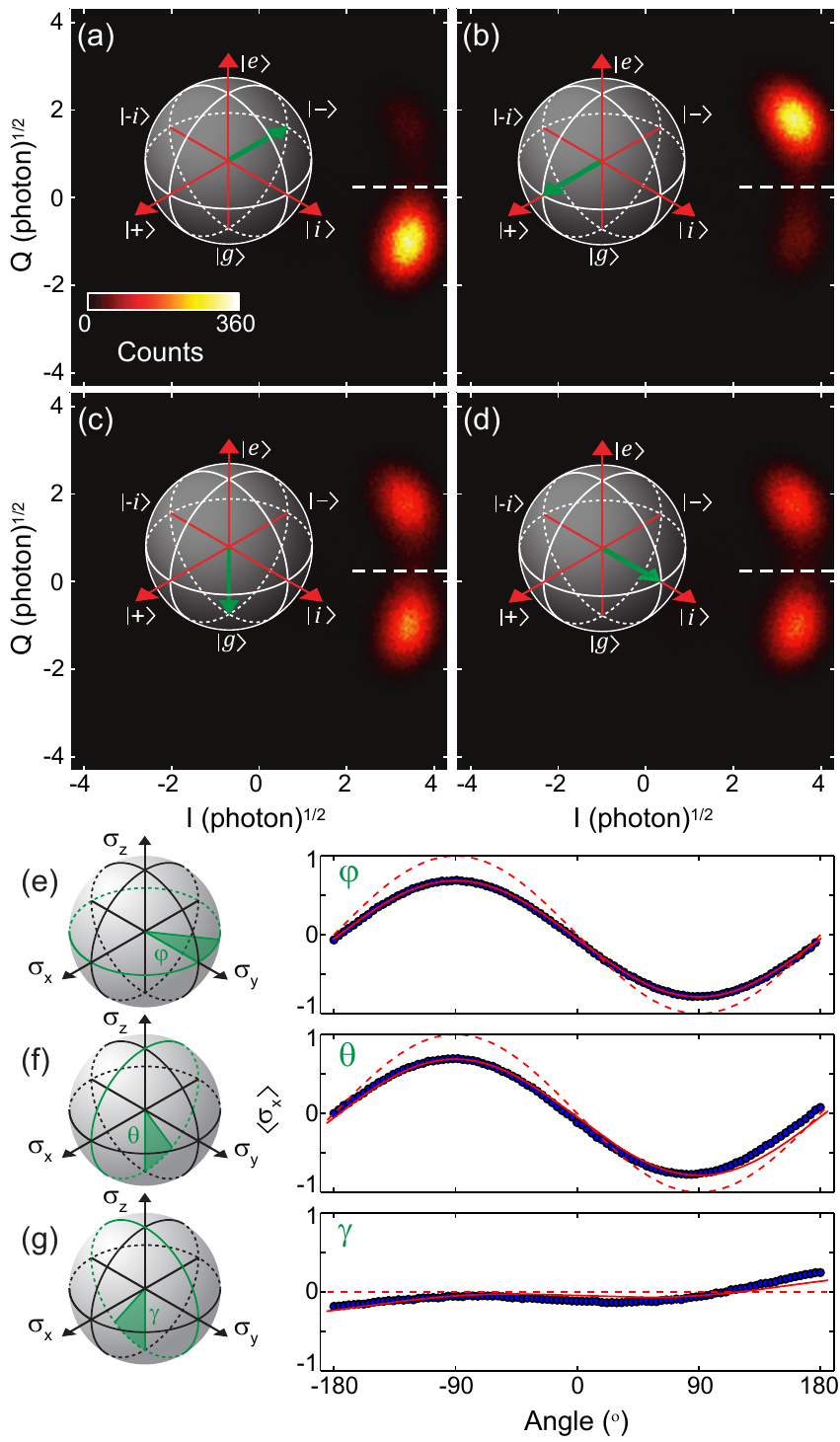}
\caption{\label{fig3}  The qubit is prepared in $\ket{-}$ (a), $\ket{+}$ (b), $\ket{g}$ (c) and $\ket{i}$ (d) as is shown by the Bloch spheres. We then measure $I,Q$ histograms in the presence of the readout, sideband and Rabi drives ($\Omega_\mathrm{R}/2\pi = 70\:\mathrm{MHz}$). The results for the initializations $\ket{-}$ and $\ket{+}$ show two Gaussian distributions separated by $5.4\:\sigma$. The initializations $\ket{g}$ and $\ket{i}$ show nearly 50:50 weight in both distributions as expected for $\left<\bm{\sigma_x}\right>$. Outcomes above (below) the separatrix (dashed white line) are identified as $\ket{+}$ ($\ket{-}$).
The expectation value $\left<\bm{\sigma_x}\right>$ is plotted in blue for states prepared on the Bloch sphere surface in the $\bm{\sigma_x}-\bm{\sigma_y}$ plane (e), $\bm{\sigma_x}-\bm{\sigma_z}$ plane (f) and $\bm{\sigma_y}-\bm{\sigma_z}$ plane (g). The dashed red lines in (e)-(g) show the ideal expectation value $\left<\bm{\sigma_x}\right>$, while the solid red lines show a theoretical prediction based on a simulation of the cavity-qubit system, including measurement imperfection.
}
\end{figure}

\begin{figure} [thb]

\includegraphics[angle = 0, width = \columnwidth]{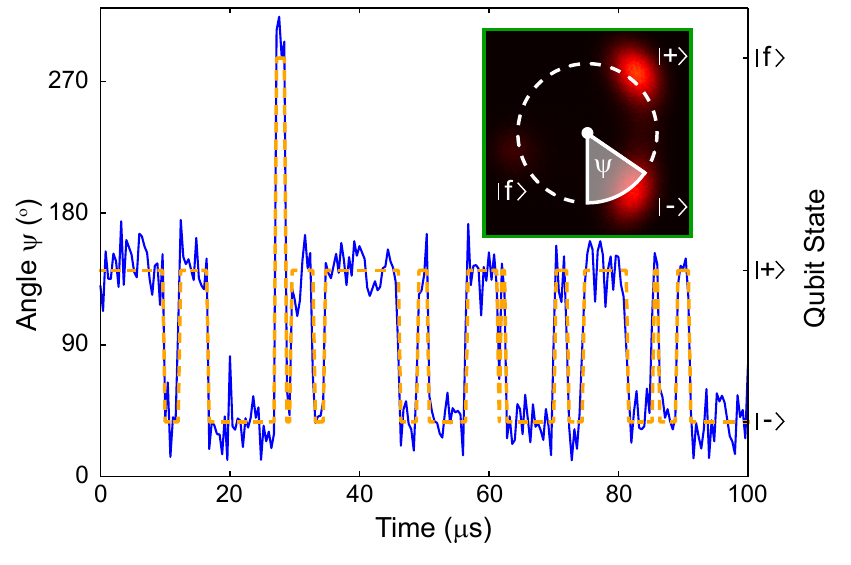}
\caption{\label{fig4}  A cut of a continuous trace of quantum jumps taken from the measurement histogrammed in Fig.~2e. The blue line shows the angle in phase space around the point shown in the inset, and we observe jumps between the eigenstates of $\bm{\sigma_x}$. The dashed orange line is a two-point filter estimate of the state of the qubit. The qubit is sometimes in its 2nd excited state ($\ket{f}$).
}
\end{figure}

Since our measurement is continuous, and is expected to be QND, we should be able to track the state of the effective $\bm{\sigma_x}$ qubit in real time. Evidence for the QND nature of our measurement is obtained from the observed quantum jumps between the states of our effective qubit. In Fig.~4, we show a cut from the $1\:\mathrm{s}$ jump trace histogrammed in Fig.~2e, where we have defined an angle $\psi$ around the circumcenter of the $\ket{-}$, $\ket{+}$ and $\ket{f}$ distributions (see inset). The dashed orange line corresponds to a two-point filter estimating the state of the qubit. The average time between jumps from $\ket{+}$ to $\ket{-}$ is $T_\mathrm{jump} = 4\:\mathrm{\mu s}$, limited by the dephasing induced by the sideband tone~\cite{Supplementary}. This induced $T_\mathrm{jump}$ limits the fidelity of our measurement, as the probability to stay in the same state during the $T_m= 400\:\mathrm{ns}$ integration length is $e^{-T_m/T_\mathrm{jump}} = 0.9$. We can also observe jumps to the qubit $\ket{f}$ state, which occur from both $\ket{-}$ and $\ket{+}$ due to the Rabi drive. In a future experiment, the dephasing induced by the photon shot-noise of the sideband~\cite{Clerk2010} could be reduced by increasing the strength of the sideband tone and its detuning from the cavity frequency (see supplementary material). We could thus observe the intrinsic quantum jumps of the effective qubit due to the inherent dephasing noise of the bare transmon, which corresponds to the effective dissipation of our $\bm{\sigma_x}$ qubit. Measuring the time correlations of quantum jumps~\cite{Vool2014} would then give access to the spectral density of qubit dephasing, potentially faster than traditional methods~\cite{Bylander2011} and including both the negative and positive frequency components~\cite{Clerk2010}.

In conclusion, we have presented a method to synthesize a tunable effective JC Hamiltonian between a cavity and an effective qubit whose eigenstates are transverse superpositions of the bare qubit. We have used this technique in the dispersive regime to observe quantum jumps between the qubit eigenstates of $\bm{\sigma_x}$ in the rotating frame. In addition to the direct application described in the previous paragraph, our experiment demonstrates a unique and simple example of the ability to engineer an effective Hamiltonian and measurement by the addition of a constant pump tone.  Such new effective quantum operations are at the heart of quantum simulation and necessary for quantum error correction.
Furthermore, our effective JC Hamiltonian could itself be tuned to reach the ultra-strong coupling regime ($g_\mathrm{eff}/\omega^\mathrm{eff}_c \approx 0.1 $) and the deep-strong coupling regime ($g_\mathrm{eff} > \omega^\mathrm{eff}_c, \omega^\mathrm{eff}_q$)~\cite{Casanova2010} which has recently been experimentally attained~\cite{Yoshihara2016,Forn-Diaz2016}. In addition, the potential sensitivity of the effect demonstrated in this paper to the effective detuning $\Delta_c - \Omega_\mathrm{R}$ could be useful for RF voltage metrology. The measurement of eigenstates of $\bm{\sigma_x}$ could also be interesting for a fundamental study of competing measurements of non-commuting variables. Recent work~\cite{Gourgy2016} has shown a protocol to measure two non-commuting Pauli operators of a qubit simultaneously. The protocol described here measures the remaining operator, and a combination of both experiments would allow us to measure all 3 Pauli operator of the qubit simultaneously with varying measurement strengths, potentially leading to novel quantum state monitoring~\cite{Ruskov2010}.

We acknowledge fruitful discussions with Shay Hacohen-Gourgy and Irfan Siddiqi.
Facilities use was supported by YINQE, the Yale SEAS cleanroom, and NSF MRSEC DMR 1119826. This research was supported by ARO under Grant No. W911NF-14-1-0011 and by MURI-ONR Grant No. 0041302 (410042-3). S.M.G. acknowledges additional support from NSF DMR-1301798.

\bibliographystyle{apsrev}
\bibliography{sigmaxref}

\end{document}


\title{Supplementary information for ``Continuous quantum nondemolition measurement of the transverse component of a qubit''}


\maketitle

\section{Derivation of the effective dispersive Hamiltonian}
We begin our treatment with the dispersive Hamiltonian of a qubit and a cavity with two pumps, one on the cavity mode with amplitude $\epsilon_{sb}$ and frequency $\omega_{sb}$ near the cavity resonance frequency $\omega_c$,
and the other on the qubit with amplitude $\Omega_\mathrm{R}$ and frequency $\omega_{qp}$ near the frequency of the qubit $\omega_q$:
\begin{equation}
\bm{H}/\hbar=\omega_c \bm{a^{\dagger}a} + \frac{\omega_q}{2} \bm{\sigma_{z}} + \frac{\chi}{2}\bm{a^{\dagger}a}\bm{\sigma_z} + \Omega_\mathrm{R} \cos(\omega_{qp} t)\bm{\sigma_x} + \epsilon_{sb} \cos (\omega_{sb} t)(\bm{a}+\bm{a^\dagger}),
\end{equation}
where $\chi$ is the dispersive shift between the qubit and the cavity. We can move to the rotating frame of both the qubit and the cavity,  $\bm{U_c}=e^{i\bm{a^{\dagger}a}\omega_{sb}t},\; \bm{U_q}=e^{i\bm{\sigma_{z}}\frac{\omega_{qp}}{2}t}$, and obtain $\bm{H} \rightarrow \bm{U_c U_q H U^\dagger_q U^\dagger_c}$
\begin{equation}
\label{RWA}
\bm{H}/\hbar=\Delta_c \bm{a^{\dagger}a} + \frac{\Delta_q}{2} \bm{\sigma_{z}} + \frac{\chi}{2}\bm{a^{\dagger}a}\bm{\sigma_z} + \frac{\Omega_\mathrm{R}}{2} \bm{\sigma_x} + \epsilon_{sb} (\bm{a}+\bm{a^\dagger)},
\end{equation}
where $\Delta_c = \omega_c-\omega_{sb}$ and $\Delta_q = \omega_q-\omega_{qp}$.
We can now eliminate the sideband pump by displacing the cavity such that its steady state, which is a coherent state, becomes the new ground state. This can be done using the displacement operator $\bm{U_d} = e^{\bar{a}^\ast \bm{a} - \bar{a} \bm{a^\dagger}}$
which is equivalent to the change of frame $\bm{a} = \bar{a} + \bm{d}$, where $\bm{d}$ is the new operator of the cavity and $\bar{a}$, a c-number, is the displacement in phase space. To eliminate the sideband pump term in Eq. \ref{RWA}, we choose $\bar{a} = \frac{-\epsilon_{sb}}{\Delta_c -i \kappa/2}$, where $\kappa$ is the energy relaxation rate of the oscillator. Under this transformation the Hamiltonian reduces to:
\begin{equation}
\label{JCham}
\bm{H}/\hbar=\Delta_c \bm{d^{\dagger}d} + \frac{\Delta_q+\bar{n}_{sb}\chi}{2} \bm{\sigma_{z}} + \frac{\Omega_\mathrm{R}}{2} \bm{\sigma_x}+ \frac{\chi}{2}(\bar{a}^\ast \bm{d} + \bar{a} \bm{d^{\dagger}} + \bm{d^\dagger d})\bm{\sigma_z},
\end{equation}
where $\bar{n}_{sb} = \bar{a}^\ast \bar{a}$. To see why the correct $\bar{a}$ needs to include the energy relaxation $\kappa$, we should look at the Lindblad decay operator ($D[\bm{L}]\rho = \bm{L}\rho\bm{L^\dagger} - \bm{L^\dagger L} \rho/2 - \rho \bm{L^\dagger L}/2$) for cavity dissipation: 
$\kappa D[\bm{a}]\rho$. Under the displacement transformation $U_d$ we get $\kappa D[\bm{a}]\rho = \kappa D[\bm{d}] \rho + \frac{\kappa}{2} [\bar{a}^\ast \bm{d} - \bar{a} \bm{d^\dagger},\rho]$, and as the last term has the form of a commutator we can add it as an effective term to the Hamiltonian in the master equation. Therefore, the effective hamiltonian for the cavity is $\Delta_c \bm{a^{\dagger}a} + \frac{i\kappa}{2}(\bar{a}^\ast \bm{d} - \bar{a} \bm{d^\dagger}) + \epsilon_d(\bm{a}+\bm{a^\dagger})$ and we can see that under the $\bm{U_d}$ transformation with the $\bar{a}$ choice mentioned above, this simply becomes $\Delta_c \bm{d^\dagger d}$ - an undriven oscillator.
\\
The next trick will be labeling $\bm{\sigma_z} = \bm{\sigma^+_x} + \bm{\sigma^-_x}$ where $\bm{\sigma^+_x},\bm{\sigma^-_x}$ are the raising and lowering operators of the $\bm{\sigma_x}$ eigenstates respectively. We can label the eigenstates of $\bm{\sigma_x}$ 
with eigenvalues $1,-1$ as $\ket{+},\ket{-}$ respectively. And so $\bm{\sigma^+_x} \ket{-} = \ket{+}$ and $\bm{\sigma^-_x} \ket{+} = \ket{-}$. With this substitution the Hamiltonian:
\begin{equation}
\bm{H}/\hbar=\Delta_c \bm{d^{\dagger}d} + \frac{\Omega_\mathrm{R}}{2} \bm{\sigma_x}+ \frac{\chi}{2}(\bar{a}^\ast \bm{d}+ \bar{a} \bm{d^{\dagger}} + \bm{d^\dagger d})(\bm{\sigma^+_x} + \bm{\sigma^-_x})
\end{equation}
resembles the Jaynes-Cummings~\cite{Jaynes1963, Haroche2006} Hamiltonian between the displaced cavity and the $\bm{\sigma_x}$ quadrature of the qubit, as discussed in the supplementary material of Ref.~\onlinecite{Murch2012}. We have assumed in the last equation that $\Delta_q = -\bar{n}_{sb}\chi$ to cancel the $\bm{\sigma_z}$ term. Note that, unlike in the main text, we do not make the rotating wave approximation and so keep the counter-rotating terms such as $\bm{d \sigma^-_x}$, as in the quantum Rabi model. These terms will add a correction to the dispersive shift, as is shown below.

If we assume that $|\Delta| = |\Omega_\mathrm{R} - \Delta_c| \gg \frac{\chi |\bar{a}|}{2}$, the coupling between the effective qubit and cavity is weak and we treat it perturbatively by performing the Schrieffer-Wolff dispersive transformation $\bm{U}=e^{\frac{\chi}{2\Delta}(\bar{a}^\ast \bm{d\sigma^+_x}-\bar{a}\bm{d^{\dagger}\sigma^-_x})}$ and keeping the terms up to first order in $\frac{\chi |\bar{a}|}{2\Delta}$ ~\cite{Schrieffer1966,Blais2004}. This transformation diagonalizes and removes the zeroth-order term $\frac{\chi}{2}(\bar{a}^\ast \bm{d\sigma^+_x}+\bar{a}\bm{d^{\dagger}\sigma^-_x})$ but we are still left with the zeroth order terms $\frac{\chi }{2}(\bar{a}^\ast \bm{d\sigma^-_x} + \bar{a} \bm{d^\dagger \sigma^+_x}) + \frac{\chi}{2}\bm{d^\dagger d}(\bm{\sigma^-_x} + \bm{\sigma^+_x})$. To see the non-rotating contribution arising from these non-stationary zeroth order terms, we can then perform similar Schrieffer-Wolff transformations $\bm{U^\prime} = e^{\frac{\chi}{2\Sigma}(\bar{a}\bm{d^\dagger \sigma^+_x}-\bar{a}^\ast \bm{d\sigma^-_x})}$ and $\bm{U^{\prime \prime}}= e^{\frac{\chi}{2\Omega_\mathrm{R}}\bm{d^\dagger d} (\bm{\sigma^+_x}-\bm{\sigma^-_x})}$ where $\Sigma = \Omega_\mathrm{R} + \Delta_c$ and we assume $|\Sigma| \gg \frac{\chi |\bar{a}|}{2}$ and $\Omega_\mathrm{R} \gg \frac{\chi}{2}$. Performing the three unitary transformations $\bm{U}$, $\bm{U^\prime}$ and $\bm{U^{\prime \prime}}$ and keeping first order terms in $\frac{\chi |\bar{a}|}{2\Delta}$, $\frac{\chi |\bar{a}|}{2\Sigma}$ and $\frac{\chi}{2\Omega_\mathrm{R}}$ we obtain the Hamiltonian:

\begin{equation}
\label{dispsigx}
\bm{H}/\hbar=\Delta_c \bm{d^{\dagger}d} + \frac{\Omega_\mathrm{R}+\zeta^\prime/2}{2} \bm{\sigma_x}+ \frac{\zeta}{2}\bm{d^\dagger d} \bm{\sigma_x} + \frac{\chi^2}{4\Omega_\mathrm{R}}\bm{d^\dagger d^\dagger d d} \bm{\sigma_x},
\end{equation}

where the Lamb shift $\zeta^\prime = \frac{\chi ^2}{2}(\frac{\bar{n}_{sb}}{\Delta} + \frac{\bar{n}_{sb}}{\Sigma})$ and the dispersive $\bm{\sigma_x}$ coupling strength $\zeta$ is equal to:

\begin{equation}
\label{zeta}
\zeta = \frac{\chi ^2}{2}(\frac{\bar{n}_{sb}}{\Delta} + \frac{\bar{n}_{sb}}{\Sigma} + \frac{1}{\Omega_\mathrm{R}}).
\end{equation}

This parameter will be the dispersive shift of the cavity frequency depending on the qubit being in the $\ket{-}$ or $\ket{+}$ state. Note that in our experiment $\Delta$, $\Sigma$ and $\Omega_\mathrm{R}$ are of the same order of magnitude so all 3 terms contribute to $\zeta$. From the histograms such as that in main text Fig.~2e, we can extract $\zeta$ as we know that when we drive the cavity with a readout drive of amplitude $\epsilon_r$ and detuning $\Delta_r$ the coherent states in the cavity corresponding to qubit states $\ket{-}$ and $\ket{+}$ are $\bar{a}_- = \frac{-\epsilon_r}{-\zeta/2 + \Delta_r -i\kappa /2}$ and $\bar{a}_+ = \frac{-\epsilon_r}{\zeta/2 + \Delta_r -i\kappa /2}$ respectively. The integrated signals we measure will then simply be $\sqrt{\kappa T_m} \bar{a}_\pm$ when $T_m$ is the integration length, $400\:\mathrm{ns}$ for our measurement. We can thus extract $\zeta = \frac{\kappa}{2}(\frac{\operatorname{Re} \bar{a}_+}{\operatorname{Im} \bar{a}_+}-\frac{\operatorname{Re} \bar{a}_-}{\operatorname{Im} \bar{a}_-})$. In supplementary Fig.~1a we show the extracted $\zeta$ values from $I, Q$ histograms taken from measurements with varying Rabi frequency $\Omega_\mathrm{R}$. The solid red line shows the theoretical prediction from Eq. \ref{zeta}. The corresponding $I, Q$ histograms can be viewed in the supplementary movie.

\begin{figure} [htb]
\includegraphics[angle = 0, width = \textwidth]{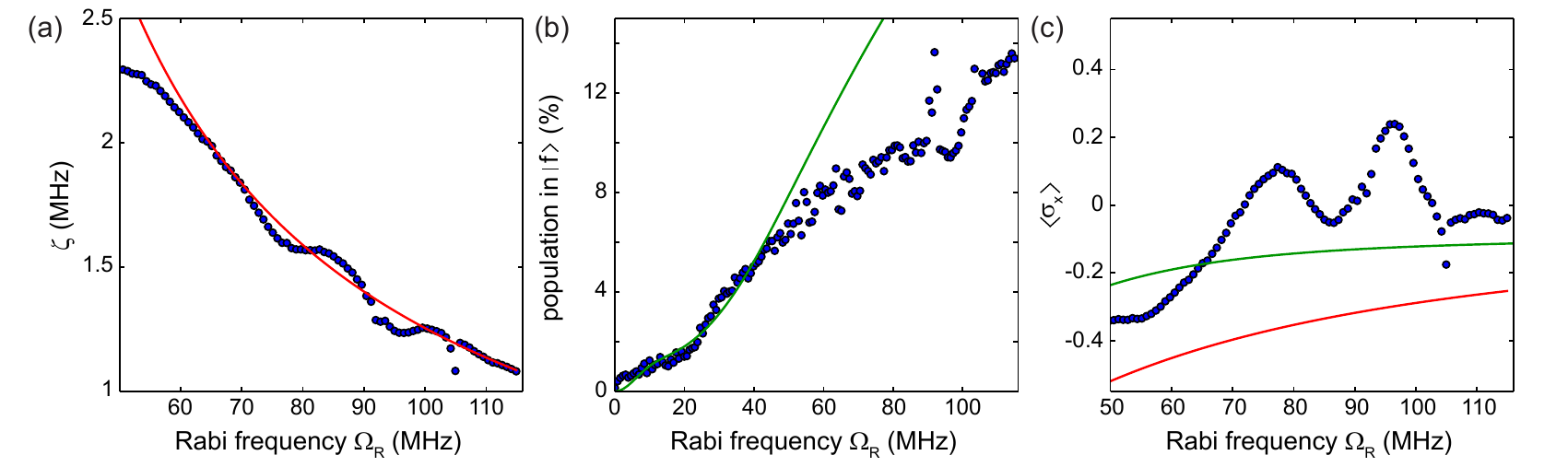}
\caption{\label{supfig1}  Parameters extracted from $I, Q$ histograms for varying Rabi frequency $\Omega_\mathrm{R}$. (a) The effective $\bm{\sigma_x}$ dispersive shift $\zeta$ vs. $\Omega_\mathrm{R}$ in the dispersive regime $|\Omega_\mathrm{R} - \Delta_c| \gg \frac{\chi |\bar{a}|}{2}$. The values were extracted from the $I,Q$ histogram as explained in the supplementary text. The theoretical prediction in Eq. \ref{zeta} is shown in red, and there is good agreement between theory and measurement. (b) The population in the 2nd excited state $\ket{f}$ vs. $\Omega_\mathrm{R}$. The green line is a numerical simulation master equation containing Hamiltonian $\bm{H} +\bm{ H_{f}}$ from Eqs. \ref{JCham} and \ref{Hf}, and decay operators. At $\Omega_\mathrm{R}/2\pi > 50\:\mathrm{MHz}$ there is a discrepancy between the theory and measurement. (c) The equilibrium expectation value $\left< \bm{\sigma_x} \right >$ vs. $\Omega_\mathrm{R}$. The red line is a theoretical prediction (see supplementary text) and the green line is a numerical simulation of a master equation containing the Hamiltonian $\bm{H} +\bm{ H_{f}}$ from Eqs. \ref{JCham} and \ref{Hf}, and decay operators. The equilibrium of the $\ket{+}$/$\ket{-}$ population does not seem to be given by a thermal bath or our predicted models, but exhibits a much more irregular behavior. }
\end{figure}
The transformation $\bm{U^{\prime \prime}}$ contributes the final term in Eq. \ref{zeta}, but it also gives rise to the final term in Eq. \ref{dispsigx} - a Kerr term on the displaced cavity dependent on the state of the $\bm{\sigma_x}$ qubit. As the readout tone we used contained $\bar{n} = 0.9$ photons, this term should not have a noticable effect on our measurement (it only affects states with 2 or more photons). With a stronger readout tone, however, one could observe an addional effect due to this cavity nonlinearity.

\section{Beyond the two-level system}
The Hamiltonian described above treats the transmon qubit as a two-level system, ignoring its higher levels. We can add a correction to the Hamiltonian due to the 2nd excited state of the qubit $\ket{f}$. The correction will be:
\begin{equation}
\label{Hf}
\bm{H_{f}}/\hbar=\alpha \bm{\ket{f}\bra{f}} + \sqrt{2}\Omega_\mathrm{R}(\bm{\ket{e}\bra{f}}+\bm{\ket{f}\bra{e}})+ \frac{3\chi}{2}(\bar{a}^\ast \bm{d }+ \bar{a} \bm{d^{\dagger}} +\bm{ d^\dagger d})\bm{\ket{f}\bra{f}}
\end{equation}
where $\alpha$ is the non-linearity of the transmon qubit. The last term shows the dispersive shift due to the qubit being in the $\ket{f}$ state. From the first two terms we can see that as the Rabi frequency increases, so does the equilibrium population in $\ket{f}$.  In supplementary Fig.~1b we show the extracted $\ket{f}$ state population as a function of the Rabi frequency $\Omega_\mathrm{R}$. The green line corresponds to a theoretical prediction based on a master equation simulation including the Hamiltonian $\bm{H}+\bm{H_{f}}$ from Eqs. \ref{JCham} and \ref{Hf} as well as decay operations for the cavity and the qubit. While the agreement is good at low Rabi frequencies, for Rabi frequency values $\Omega_\mathrm{R}/2\pi > 50\:\mathrm{MHz}$, the measured $\ket{f}$ state population is significantly lower than its predicted value. 

We verified this effect is not due to even higher excited states by simulating the qubit as 7-level system with fourth order nonlinearity and found similiar results. The $\ket{f}$ behavior was unchanged and the 3rd excited state was populated up to $4\%$ for $\Omega_\mathrm{R}/2\pi = 100\:\mathrm{MHz}$. All higher states were never populated. We currently do not understand the reason for the discrepancy between the measured and predicted population of the $\ket{f}$ for high values of $\Omega_\mathrm{R}$.

\section{Decay and thermalization}
In addition to the inherent lifetime of the qubit, the dispersive coupling between the qubit and the cavity induces a dephasing on the qubit if the cavity is populated with photons~\cite{Gambetta2006,Clerk2010}. For a coherent tone applied on the cavity (in our case the sideband tone), we can calculate the photon-photon correlation function and obtain the cavity photon shot noise spectral density $S_{nn}[\omega] = \bar{n}_{sb} \kappa/((\omega-\Delta_c)^2 + (\kappa/2)^2$). The presence of the dispersive coupling term $\frac{\chi}{2} \bm{a^\dagger a \sigma_z}$ induces a dephasing of the qubit, which can be calculated using Fermi's golden rule as $\Gamma_{+-} = (\chi/2)^2 S_{nn}[\omega]$, where $\Gamma_{+-}$ is the transition rate from $\ket{+}$ to $\ket{-}$ and $\omega$  is the energy splitting between these two levels - in our case equal to $\Omega_\mathrm{R}$. The opposite transition rate $\Gamma_{-+}$ is proportional to the spectral density with the negative frequency $-\Omega_\mathrm{R}$. There is a correction to the equation above~\cite{Gambetta2006} due to $\chi \sim \kappa$ which we will neglect in this treatment for simplicity. \\
Notice that for the dispersive case $|\Delta| \gg \frac{\chi |\bar{a}|}{2}$, the dispersive transformations described above ($\bm{U}$ and $\bm{U^\prime}$) mix the qubit and cavity operators such that $\bm{d} \simeq  \bm{d} +\frac{\chi \bar{a}}{2 \Delta} \bm{\sigma^-_x} + \frac{\chi \bar{a}}{2 \Sigma} \bm{\sigma^+_x}$. The cavity decay $\kappa$ thus induces a qubit $\bm{\sigma_x}$ decay $\Gamma_{+-} = \kappa (\frac{\chi |\bar{a}|}{2 \Delta})^2$, $\Gamma_{-+} = \kappa (\frac{\chi |\bar{a}|}{2 \Sigma})^2$ which is exactly equal to the measurement-induced dephasing result for $\Delta, \Sigma \gg \kappa$. Thus we can interpret this dephasing as a decay of the effective qubit due to the Purcell effect.

We can compare this theoretical model to our measurements. For $\Omega_\mathrm{R}/2\pi = 70.5\: \mathrm{MHz}$ we can get $1/\Gamma_{+-} = 4\:\mathrm{\mu s}$ which agrees the value measured using quantum jumps (see main text Fig.~4). However, the theory predicts $1/\Gamma_{-+} = 9\:\mathrm{\mu s}$, while our measurement shows it to be $4\:\mathrm{\mu s}$ as well, leading to equal population in $\ket{+}$ and $\ket{-}$. This surprisingly high temperature of the effective qubit is currently not understood, and additionally for different values of $\Omega_\mathrm{R}$ the behavior is even more interesting. In supplementary Fig.~1c we show the equilibrium expectation value $\left< \bm{\sigma_x} \right >$ extracted from the population in the $\ket{+}$ and $\ket{-}$ distributions of the $I, Q$ histogram, for different values $\Omega_\mathrm{R}$ in the dispersive regime. The population fluctuates as a function of the Rabi frequency, sometimes with more population in $\ket{+}$ and sometimes in $\ket{-}$. In red we plot the theoretical prediction from the calculation presented above. The green line is the result of a master equation simulation of the Hamiltonian $\bm{H}+\bm{H_{f}}$ from Eqs. \ref{JCham} and \ref{Hf} and decay operators for the cavity and qubit, which includes the effect due to the 2nd excited state $\ket{f}$. The inclusion of the $\ket{f}$ state does predict a higher population in the $\ket{+}$ state as is shown, but it still does not explain the dynamics of the equilibrium expectation value as a function of the Rabi frequency.

Note that the dispersive shift $\zeta$ scales like $\frac{\bar{n}_{sb}}{(\Omega_\mathrm{R}-\Delta_c)}$ while the dephasing scales like $\frac{\bar{n}_{sb}}{(\Omega_\mathrm{R}-\Delta_c)^2}$. This distinction could allow us to increase both the cavity pump power and its detuning, thus keeping the dispersive shift constant while significantly reducing the induced dephasing - allowing us to observe quantum jumps due to the intrinstic qubit dephasing.

\section{Master equation simulation for main text Fig.~3e-g}
In main text Fig.~3e-g we show the value of $\left< \bm{\sigma_x} \right >$ extracted from measured histograms for a qubit prepared along the 3 main axes of the Bloch sphere. To understand the behavior of the system we can simulate the master equation containing the Hamiltonian $\bm{H}+\bm{H_{f}}$ from Eqs. \ref{JCham} and \ref{Hf} and the decay terms $\kappa$, $T_1$ and $T_2$. In our case $\Omega_\mathrm{R}$ is a time varying term, shaped like a Gaussian with a $4\:\mathrm{ns}$ $\sigma$ width to simulate the physical pulse that we applied. The simulations were done using the QuTiP  toolbox in python~\cite{Johansson2012}.
Note that the sideband tone was on during the pulse preparation because the qubit frame needed to be preserved, the tone induces a $\chi \bar{n}_{sb}$ qubit frequency shift and we need to work in the correct qubit frame. With the sideband tone on and qubit tone off ($\Omega_\mathrm{R} = 0$), the sideband-induced lifetime of the qubit (see previous section) is $T_2 = 2/(\chi^2 S_{nn}[\omega=0])= 150\:\mathrm{ns}$ which clearly limits the fidelity of our preparation pulses. This also agrees with independent $T_2$ measurements in the presence of the sideband tone but not the qubit tone.

The simulation results accounted for the errors during state preparation. Additionally, our measurement is imperfect as the state can jump during the measurement ($e^{-T_m/T_{jump}} = 90\%$), or be assigned incorrectly due to overlap in the Gaussian distributions ($1\%$). We can combine these two effects to predict $89\%$ measurement fidelity, while we experimentally obtain $88\%$ by measuring twice in a row and seeing the probability of agreement. There is an additional $2\%$ shift as our measurement slightly prefers the $\ket{-}$ state to the $\ket{+}$ state.

\bibliographystyle{apsrev}

\bibliography{sigmaxref}